\pdfoutput=1
\documentclass[11pt]{article}
\usepackage[final]{acl}

\usepackage{times}
\usepackage{latexsym}
\usepackage[T1]{fontenc}
\usepackage[utf8]{inputenc}
\usepackage{microtype}
\usepackage{inconsolata}
\usepackage{graphicx}

\usepackage{enumitem}
\usepackage{amsmath}
\usepackage{amssymb}
\usepackage{multirow}
\usepackage{booktabs}
\usepackage{bibentry}
\usepackage{tabularray}
\usepackage{graphicx}
\usepackage{svg}
\usepackage{tabularx}

\usepackage{tikz}
\usepackage{xcolor}
\usepackage{colortbl} % 如果你使用了 \colorbox
\newcolumntype{L}{>{\raggedright\arraybackslash}X}

\definecolor{customborder}{RGB}{206,157,53}
\newcommand{\redDashedBox}[1]{%
  \tikz[baseline=(word.base)] \node[
    draw=red, 
    solid, 
    line width=1.5pt,
    inner sep=1.5pt,
    rounded corners=2pt
  ] (word) {#1};%
}

\title{Chain-Talker: Chain Understanding and Rendering for \\ Empathetic Conversational Speech Synthesis}

\author{
  Yifan Hu$^1$, Rui Liu$^{1}$\thanks{Corresponding author.}, Yi Ren$^2$, Xiang Yin$^2$, Haizhou Li$^{3}$ \\
  $^1$~Inner Mongolia University, Hohhot, China \\
  $^2$~ByteDance, Singapore \\
  $^3$~SRIBD, School of Data Science, The Chinese University of Hong Kong, Shenzhen, China \\
  \texttt{22309013@mail.imu.edu.cn, imucslr@imu.edu.cn},\\ 
  \texttt{\{ren.yi, yinxiang.stephen\} @bytedance.com, haizhouli@cuhk.edu.cn}
}

\begin{document}

\maketitle
\begin{abstract}
Conversational Speech Synthesis (CSS) aims to align synthesized speech with the emotional and stylistic context of user-agent interactions to achieve empathy.
Current generative CSS models face interpretability limitations due to insufficient emotional perception and redundant discrete speech coding.
To address the above issues, we present \textbf{Chain-Talker}, a three-stage framework mimicking human cognition: {\textit{Emotion Understanding}} derives context-aware emotion descriptors from dialogue history; {\textit{Semantic Understanding}} generates compact semantic codes via serialized prediction; and {\textit{Empathetic Rendering}} synthesizes expressive speech by integrating both components. To support emotion modeling, we develop \textbf{CSS-EmCap}, an LLM-driven automated pipeline for generating precise conversational speech emotion captions. Experiments on three benchmark datasets demonstrate that Chain-Talker produces more expressive and empathetic speech than existing methods, with CSS-EmCap contributing to reliable emotion modeling. \textcolor[rgb]{0.93,0.0,0.47}{The code and demos are available at: \url{https://github.com/AI-S2-Lab/Chain-Talker}.}

\end{abstract}

\section{Introduction}
Conversational speech synthesis (CSS) aims to express a target utterance with the proper linguistic and affective prosody in a user-agent conversational context \cite{guo2021conversational}. This task not only requires the agent to accurately perceive the user's emotion but also to ensure that the generated speech's emotion and style align with the conversational situation. In recent years, with the development of human-computer interaction (HCI), CSS has become an integral part of intelligent interactive systems and plays an important role in areas such as virtual assistants \cite{jain2024spear} and voice agents \cite{jaber2024cooking}.

\begin{figure}[t]
\centering
\centerline{
\includegraphics[width=1\linewidth]{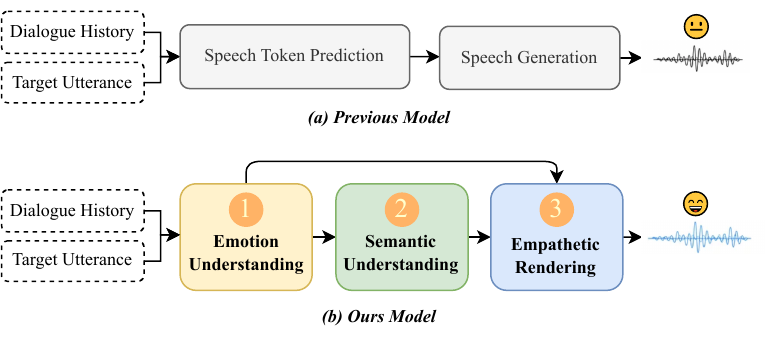}
}
\caption{(a) Previous methods predict speech tokens directly based on context. (b) Our approach progressively realizes empathetic CSS through three stages: Emotion Understanding, Semantic Understanding, and Empathetic Rendering.}
\label{fig:case}
\end{figure}

Traditional CSS attempts mainly focus on taking the multi-modal dialogue history, including the text and speech modalities, to predict the speech representations of the speech to be synthesized. Afterward, these representations are fed to the speech synthesizer to decode the target conversational speech. 
% \textcolor{red}{list 2 -3 works, such as ECSS.} 
In this process, elaborate encoding modules were introduced to enhance speech quality by incorporating style embeddings \cite{guo2021conversational, Yuto2022Acoustic, xue2023M2ctts} or emotional category information \cite{liuYuchen2023Emotionally, dengYaYue2023CMCU-CSS, Liu2024ECSS} into the representations. 
% Although naturalness increased, the expressiveness and robustness of the synthesized speech remain limited.
Recently, advanced CSS models like GPT-Talker \cite{liu2024generative}, based on Generative Pre-trained Transformer (GPT) \cite{radford2018improving}, have significantly enhanced the naturalness and expressiveness of synthesized speech by directly predicting speech token sequences (such as HuBERT encoding \cite{Hsu2021hubert}) from dialogue contexts, as shown in Fig.~\ref{fig:case} (a). Such a process lacks interpretability in two ways: 1) Speech generation does not fully understand the emotion of the conversation, making it difficult to achieve true empathy.
\textcolor{black}{However, using natural language descriptions allows easier control and representation of style and emotion in speech \cite{Guo2023prompttts, ji2024textrolspeech}. This approach directly establishes a strong correlation between the semantic content and the acoustic expressiveness \cite{Yang2024InstructTTS}. Therefore, understanding captions enables the comprehension of emotional changes in the dialogue.}
% \textcolor{red}{list some works about speech emotion captions to explain caption is a good way to understand the emotion and it has a \textbf{strong relation} with the semantic.}
2) General discrete speech codes contain too much redundant information. They are often obtained by quantizing intermediate representations from pre-trained models \cite{Hsu2021hubert} or using Neural Audio Codec models \cite{meil2022soundstream}, which mix semantic and acoustic information and have limited expressive capacity.

% For example, some studies primarily explore multi-scale semantic and acoustic dependencies within dialogue contexts \cite{guo2021conversational, huyifan2022FCTalker, Yuto2022Acoustic, xue2023M2ctts} or employ self-supervised \cite{DengYayue2023CONCSS} and supervised \cite{liuYuchen2023Emotionally} learning methods to predict continuous style representations of target utterances, aiming to achieve dialog-aware conversational speech synthesis.
% To further enhance the expressiveness of the target speech, CMCU-CSS \citet{dengYaYue2023CMCU-CSS} and ECSS \citet{Liu2024ECSS} also predict the emotional category of the target speech directly, allowing the target speech to contain more emotional information.

% However, these methods involve complex model architectures, require careful manual fine-tuning, and lack robustness. Recently, advanced CSS models like GPT-Talker \cite{liu2024generative}, based on Generative Pre-trained Transformer (GPT) \cite{radford2018improving}, have significantly enhanced the naturalness and expressiveness of synthesized speech by directly predicting speech token sequences from dialogue contexts. Nevertheless, speech token sequences are often obtained by quantizing intermediate representations from pre-trained models \cite{Hsu2021hubert} or using Neural Audio Codec models \cite{meil2022soundstream}, which mix semantic and acoustic information and have limited expressive capacity. This makes it difficult for models to accurately perceive and understand emotional changes in conversations.

To address the above issues, we propose a Chain Understanding and Rendering scheme for Empathetic CSS, termed \textbf{Chain-Talker}.
Drawing on the chain-like human thinking process \cite{zheng2023ddcot, ShimaD2023MathPrompter, Huang2024ECRChain}, Chain-Talker decomposes CSS into a three-link thinking chain including \textit{Emotion Understanding}, \textit{Semantic Understanding}, and \textit{Empathetic Rendering}.
As shown in the Fig.~\ref{fig:case} (b), the emotion understanding module perceives the emotion description of the current utterance based on the conversation history with the conversation-related speech emotion description. The semantic understanding module continues to generate purely semantic codes of the speech by means of serialization prediction, and then the emotion description and semantic codes are used for the final empathetic CSS.
This cognitive chain architecture ensures precise comprehension of contextual emotional states, enabling accurate affective responses in human-machine dialogues. By decoupling emotion and semantics into modularized processes (emotion understanding and semantic understanding), the system achieves independent yet synergistic modeling, forming an interpretable empathetic CSS framework with transparent decision-making mechanisms.

To ensure that Chain-Talker learns a robust understanding of expressiveness such as emotion and style during training, we propose an LLM-driven automatic dialog-aware empathetic captioning pipeline, \textbf{CSS-EmCap}, for conversational speech. 
We employ the CSS-EmCap pipeline to generate emotional descriptions for three benchmarking CSS datasets, including NCSSD \cite{liu2024generative}, MultiDialog \cite{se2024multidialog} and DailyTalk \cite{lee2023dailytalk}. Three datasets with emotional descriptive information are consolidated as the final training data for Chain-Talker.
Subjective and objective experiments are conducted to verify the reliability of the proposed pipeline and the effectiveness of Chain-Talker. The results indicate that Chain-Talker outperforms other CSS baseline models by synthesizing more appropriate and empathetic conversational speech, highlighting the necessity of the proposed pipeline.
\textcolor{black}{
In summary, the main contributions of this paper are:
\begin{itemize}
% [leftmargin=8pt]
 \item We introduce Chain-Talker, which employs a three-stage chain modeling process. After perceiving the emotions in the dialogue and serially generating semantic codes, it collaboratively produces empathetic response speech.
 \item We propose CSS-EmCap, an LLM-driven automatic dialog-aware empathetic captions annotation pipeline for dialogue speech. A total of approximately 384 hours across three benchmarking CSS datasets were annotated.
 \item Comprehensive experiments prove the reliability of CSS-EmCap and the effectiveness of Chain-Talker.
 % We conduct comprehensive experiments to verify the reliability of the CSS-EmCap and the effectiveness of the Chain-Talker. Both subjective and objective experiments demonstrate that our model can reason and render more empathetic conversational speech compared to all advanced CSS baselines.
\end{itemize}
}

% //////////////////////////////////////

\begin{figure*}[t]
\centering
\centerline{
\includegraphics[width=\linewidth]{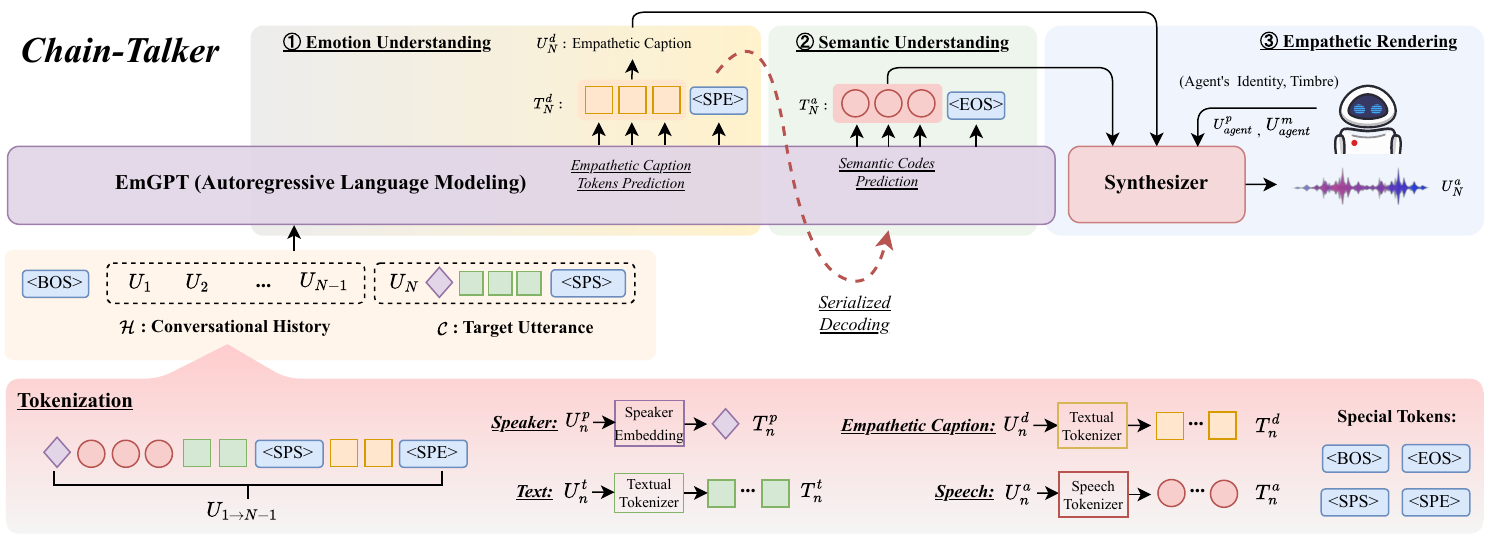}
}
\caption{The overall architecture of Chain-Talker. Chain-Talker comprises two main components: EmGPT and Synthesizer. EmGPT is responsible for emotion and semantic understanding, while the Synthesizer handles the generation of empathetic speech rendering. 
% The tokenization process involves processing various modalities of the dialogue context, which are then concatenated in a specified sequence to serve as input for EmGPT.
}
% \caption{Architecture Overview. This work comprises two core components: (1) \textbf{Speech-Instruct Alignment}, using CSS-Instruct pipeline. The dialogue data annotated with empathetic instructions generated by this pipeline is used to train the E$^{2}$-CSS model. (2) \textbf{Chain-of-Thought Instruction Reasoning and Expressive Speech Response}, the E$^{2}$-CSS model first infers the empathetic instruction $X^C_N$ for the target utterance based on the conversational context (including $Speaker: X^P_i, Text: X^T_i, Instruct: X^C_i, Speech: X^S_i $), then sequentially predicts the speech token sequence $D^S_N$, and finally, under the guidance of the empathetic instruction and in combination with the speech token sequence and information such as the agent's timbre $X^S_{agent}$ and identity $X^P_{agent}$, achieves empathetic speech $X^S_N$ rendering and response.}
\label{fig:model}
\end{figure*}

\section{Related Works}

\subsection{Chain Modeling in Conversation}
% \textcolor{red}{list some chain works, then highlight we are the first to introduce chain into the CSS task.}
\textcolor{black}{
Recently, using chain modeling to solve complex problems step-by-step has been very successful in dialogue-related tasks.  For instance, in generating text responses, \citet{chen2021reasoning} and \citet{lin2024paralingpt}  incorporate reading comprehension and sentiment analysis, enabling models to focus on key information and emotional cues, thereby improving response accuracy and relevance. In speech response tasks, systems like USDM \cite{kim2024USDM} and Spectron \cite{nachmani2024spoken} perform speech recognition before generating responses. This sequential approach within a unified LLM reduces errors from multi-module setups and enhances semantic coherence. Drawing inspiration from these successes, we pioneer the application of chain modeling to CSS tasks. This approach enables more empathetic conversations by step-by-step understanding context and rendering.
}

\subsection{Speech Emotion Description}
% \textcolor{red}{list some Speech Emotion caption works, then highlight we are different. 1) we focus on conversational speech not single sentence. 2) we use various pretrained LLM to form an LLM-based scehme.}
\textcolor{black}{
Accurately describing emotion and style in speech using natural language becomes a key research area. Traditionally, this task relies heavily on manual annotation, where annotators write adjectives or sentences based on the speech. This method is inefficient, costly, and the quality of annotations declines with prolonged manual labeling \cite{Yang2024InstructTTS, Liu2023PromptStyle, Kawamura24LibriTTSP}. 
To address these issues, some studies rely on limited manual annotations and expand the data using pre-trained models such as SimBERT or GPT-3.5 Turbo \cite{Guo2023prompttts, ji2024textrolspeech}. However, these models lack genuine speech understanding and merely rewrite sentences based on keywords, often resulting in inaccurate descriptions of emotion or speaking style. Other approaches \cite{chu2024qwen2-audio, wu2024secap,lian2024ovmer} attempt to train models capable of labeling speech style, but they typically overlook the need for emotional understanding in conversational contexts.
% To address these issues, some studies use limited manual annotations and expand them with pre-trained models like sinBERT or GPT-3.5 Turbo \cite{Guo2023prompttts, ji2024textrolspeech}. However, these models lack true speech understanding and merely rewrite sentences using keywords, often resulting in inaccurate emotional or stylistic descriptions. 
% Moreover, these methods ignore speech emotion understanding in dialogue scenarios.
To overcome these issues, we develop an automated, dialog-aware empathetic description process using LLMs.
Specifically, the proposed method first extracts stylistic attributes at the sentence-level and emotions at the dialog-level. It then prompts the LLM to generate basic descriptions, which are subsequently optimized into empathetic captions. Throughout the process, it continually prompts the LLM to reference diverse contextual information to ensure consistency between descriptions and speech.
}

\subsection{Speech Discrete Encoding}

% \textcolor{red}{list some Speech encoding works, then highlight we select the pure semantic tokens, that are optimal for our Chain model.}
\textcolor{black}{Speech tokens extracted through unsupervised learning \cite{Hsu2021hubert, meil2022soundstream} enable the synthesis of relatively natural-sounding speech \cite{gptsovits, Wang2023valle}. However, training language models with these tokens often results in slow convergence and poor stability. To address this, research has shown that semantic tokens derived from supervised learning—by capturing clear semantic information in speech and aligning it with text—can improve model stability \cite{du2024cosyvoice}. In CSS tasks, the input sequence includes various modalities, making the modeling process challenging unless it is performed within the same semantic space. Therefore, following CosyVoice, we employ a supervised automatic speech recognition (ASR) model to create a supervised semantic speech tokenizer.}

\section{Task Definition}
In user-agent spoken dialogue interactions, the user initiates the conversation, followed by the agent responding within the context of the dialogue. As the conversation progresses with alternating turns, the accumulated spoken content forms the dialogue history. The current dialogue turn is denoted by $N$, the utterance to be synthesized is represented as $\mathcal{C} = (U_N^p, U_N^t)$, and the dialogue history can be defined as 
% $\mathcal{H}$ = $\{(U_1^p, U_1^a, U_1^t, U_1^d), (U_2^p, U_2^a, U_2^t, U_2^d), \ldots, (U_{N-1}^p, \\ U_{N-1}^a, U_{N-1}^t, U_{N-1}^d,)\}$. 
$\mathcal{H}$ = $\{U_1^{p,a,t,d}, U_2^{p,a,t,d}, \ldots, U_{N-1}^{p,a,t,d}\}$. 
Here, $U_n^p$, $U_n^a$, $U_n^t$, and $U_n^d$ represent the speaker, speech, text, and emotional description of the utterance at the $n$-th turn, respectively. To generate empathetic speech, the Chain-Talker first determines the empathetic caption $U_N^d$ based on the dialogue history $\mathcal{H}$ and the current utterance $\mathcal{C}$. Next, it serially generates the semantic codes $T_{N}^a$, and finally synthesizes the corresponding expressive speech $U_N^a$ that aligns with the inferred empathetic caption.

% with the agent's identity $U_{agent}^p$ and timbre $U_{agent}^m$.

\section{Methodlogy: Chain-Talker}
This section offers a detailed description of our proposed Chain-Talker. We begin by outlining the data input format for the model, referred to as the \textit{Unified Context Tokenization} method. Building on the chain modeling, our approach has three main processes: 1) \textit{Emotion Understanding}, which focuses on predicting empathetic captions, 2) \textit{Semantic Understanding}, which aims at inferring speech codes that include semantic information, and 3) \textit{Empathetic Rendering},  which synthesizes the empathetic response speech. Finally, we conclude by discussing the model's training strategy: \textit{Multi-Stage Training}.

\subsection{Unified Context Tokenization}
To better model the context of dialogues, we follow the method of GPT-Talker, which involves the alternating concatenation of user and agent utterances to simulate a real dialogue flow.
In particular, each utterance is concatenated in the order of speaker information, speech, textual content, and empathetic captions. This allows the model to first understand the context and predict the emotion before generating the corresponding speech. Therefore, the input to Chain-Talker can be represented as $\mathcal{Q}$: 
\begin{equation}
 \mathcal{Q} = (\langle \text{BOS} \rangle, \mathcal{H}, \mathcal{C}, \langle \text{EOS} \rangle)
\end{equation}
where $\mathcal{H}$ denotes the dialogue history, consisting of $N-1$ sextuples $(U_n^p, U_n^a, U_n^t, \langle \text{SPS} \rangle, U_n^d, \langle \text{SPE} \rangle)$. $\mathcal{C}$ represents the target utterance to be synthesized, comprising a triple $(U_n^p, U_n^t, \langle \text{SPS} \rangle)$. The special tokens \cite{zhang2023speechgpt} \(\langle \text{BOS} \rangle\) and \(\langle \text{EOS} \rangle\) indicate the start and end of the entire input sequence $\mathcal{Q}$, respectively, while \(\langle \text{SPS} \rangle\) and \(\langle \text{SPE} \rangle\) mark the beginning and end of the empathetic captions \( U_n^d \).

To better represent the different modal information in the input sequence, we encode the textual content as \( T_n^t \) and the empathetic caption as \( T_n^d \) using Byte Pair Encoding (BPE) \cite{gage1994BPE}. Following CosyVoice, we extract the speaker vectors as \( T_n^p \) using a pre-trained voice-print model \footnote{https://github.com/alibaba-damo-academy/3D-Speaker/tree/main/egs/3dspeaker/sv-cam++}. Additionally, we employ a supervised automatic speech recognition model with an inserted vector quantizer (VQ) \cite{Gao2023funasr} to encode the speech as \( T_n^a \).

% To better represent the different modal information in the input sequence, we conduct the following encoding and discretization processes:
% 1) \textbf{Speaker Information ($U_n^p$)}, we extract speaker embedding vectors $T_n^p$ from each utterance's speech $U_n^a$ using a pre-trained voice-print model \footnote{https://github.com/alibaba-damo-academy/3D-Speaker/tree/main/egs/3dspeaker/sv-cam++}.
% 2) \textbf{Speech ($U_n^a$)}, following CosyVoice, we employ a supervised semantic speech tokenizer to discretize the speech. Specifically, a vector quantizer (VQ) was integrated into a supervised automatic speech recognition model \cite{Gao2023funasr} to extract discrete tokens $T_n^a$. This preserves rich semantic content information and helps LLM better model the context.
% 3) \textbf{Textual Content ($U_n^t$) and Empathetic Caption ($U_n^d$)}, we employ a Byte Pair Encoding (BPE) \cite{gage1994BPE} tokenizer to extract text tokens $T_n^t$ and caption tokens $T_n^d$.

\subsection{Emotion Understanding}
In the ``\normalsize{\textcircled{\scriptsize{1}}}\normalsize'' part of Fig. \ref{fig:model}, we utilize the dialogue context $\mathcal{Q}$ ($\mathcal{H}$ and $\mathcal{C}$) as a prompt and apply the autoregressive EmGPT to comprehend the interactions between the user and the agent, as well as the changes in emotional states within the context. Simultaneously, EmGPT predicts the empathetic caption tokens $T_N^d$ of the target utterance until the $\langle SPE \rangle$ token is predicted:
\begin{equation}
\resizebox{0.8\columnwidth}{!}{$
\begin{array}{c}
p(T_{N,:}^d| \Re_{1\rightarrow N-1}, T_{N,:}^p,T_{N,:}^t; \Theta) 
\\=
\displaystyle\prod_{j=0}^{D}p(T_{N,j}^d|T_{N,<j}^d, \Re_{1\rightarrow N-1}, T_{N,:}^p,T_{N,:}^t; \Theta)
\end{array}
$}
\end{equation}
where, $\Theta$ represents EmGPT, $\Re_{1 \rightarrow N-1}$ is the set $\{\, (T_n^p, T_n^a, T_n^t, T_n^d) \,\}_{1 \rightarrow N-1}$, $j$ denotes the value of the $j$-th token of $T_N^d$, and $D$ is the length of $T_N^d$.

\subsection{Semantic Understanding}
In the ``\normalsize{\textcircled{\scriptsize{2}}}\normalsize'' part of Fig. \ref{fig:model}, once EmGPT comprehends the context and predicts an appropriate empathetic caption $U_N^d$, it utilizes this information in conjunction with the contextual content 
$\mathcal{Q}$ to further predict the speech codes $T_N^a$ that contain the semantic information of the target utterance:
\begin{equation}
\resizebox{0.8\columnwidth}{!}{$
\begin{array}{c}
p(T_{N,:}^a|\Re_{1 \rightarrow N-1}, T_{N}^p,T_{N}^t,T_{N}^d;\Theta)
\\ = 
p(T_{N,:}^d|\Re_{1 \rightarrow N-1}, T_{N}^p, T_{N}^t;\Theta) 
\\ \cdot 
\displaystyle\prod_{i=0}^{A}p(T_{N,i}^a|T_{N,<i}^a, \Re_{1 \rightarrow N-1}, T_{N}^p,T_{N}^t;\Theta)
\end{array}
$}
\end{equation}
where $i$ denotes the value of the $i$-th token of $T_N^a$, and $A$ is the length of $T_N^a$.

During the training EmGPT phase, we use a teacher forcing approach where the left-shifted sequence serves as the input pattern and the original sequence acts as the target output. We divide the loss function into two components: \( \mathcal{L}_{\text{caption}} \) and \( \mathcal{L}_{\text{speech}} \). They compute the cross-entropy loss between the true and predicted values for empathetic caption tokens and semantic codes, respectively.

\subsection{Empathetic Rendering}
In the ``\normalsize{\textcircled{\scriptsize{3}}}\normalsize'' part of Fig. \ref{fig:model}, unlike traditional models that directly decode speech tokens into speech responses, our approach uses previously predicted empathetic captions to guide emotion and style rendering during decoding. 
% This helps the Synthesizer generate more appropriate and suitable speech. 
Specifically, the Synthesizer employs an optimal-transport conditional flow matching model (OT-CFM) \cite{du2024cosyvoice} as its backbone to predict Mel spectrograms and uses HIFI-GAN vocoder \cite{kong2020hifigan} to synthesize the waveform.

% Specifically, OT-CFM utilizes Continuous-time Normalizing Flows (CNFs) by defining a time-dependent vector field \( \nu_t(X) \) to gradually transform the prior distribution \( p_0(X) \) (set as a standard normal distribution) into the data distribution \( q(X) \) (the actual Mel spectrogram distribution). 
To enhance the quality and consistency of the generated speech, OT-CFM relies not only on the Mel spectrogram \( X \) and time step \( t \), but also incorporates empathetic captions $U_N^d$, agent's speaker information $U_{agent}^p$, semantic codes $T_N^a$, and agent's masked Mel spectrograms $U_{agent}^m$ into the prediction of the vector field. This process is specifically represented by the following differential equation:
\begin{equation}
\scalebox{0.9}{$
  \begin{aligned}
    \frac{d\phi_t(X)}{dt} = \nu_t(\phi_t(X), t \mid U_{agent}^p, U_N^d, T_N^a, U_{agent}^m)
    \end{aligned}
$}
\end{equation}
where $t \in [0, 1]$, 
% \( \phi_0(X) \sim p_0(X) \) indicates that the samples at the initial time step are drawn from the prior distribution, and \( \phi_1(X) \approx q(X) \) signifies that the samples approximate the data distribution after the flow transformation.
empathetic captions are encoded using a pre-trained sentence-level BERT model \footnote{https://huggingface.co/sentence-transformers/distiluse-base-multilingual-cased-v1} and integrated with each semantic code. Other settings follow CosyVoice.

To ensure that the model learns the correct vector field \( v_t(X) \), OT-CFM introduces Optimal Transport (OT) flows and trains the model by minimizing the difference between the predicted vector field and the theoretical OT vector field. The loss function is defined as:
\begin{equation}
\scalebox{0.85}{$
  \begin{aligned}
    \mathcal{L}_{\text{OT-CFM}} &= \mathbb{E}_{t, X_0, X_1} \left[ \left\| \omega_t\left(\phi_t^{\text{OT}}(X_0, X_1) \mid X_1\right) \right. \right. \\
    &\quad \left. \left. - \nu_t\left(\phi_t^{\text{OT}}(X_0, X_1) \mid \theta\right) \right\| \right]
  \end{aligned}
$}
\end{equation}
where $\phi_t^{\text{OT}}(X_0, X_1)$ represents the optimal transport flow, i.e., the path from $X_0$ to $X_1$. $\omega_t(\phi_t^{\text{OT}}(X_0, X_1) | X_1) = X_1 - (1-\sigma)X_0$. $\theta$ is the parameter of the neural network, used to predict the vector field $\nu_t$.

\subsection{Multi-Stage Training}
In the Chain-Talker framework, the training of EmGPT is divided into two stages: 
1) \textbf{First-Stage}: The model is trained using single-sentence text-to-speech pair data, which equips the model with the basic capability to generate speech from text. In this work, we use ``CosyVoice-300M-25Hz'' \footnote{https://www.modelscope.cn/models/iic/CosyVoice-300M-25Hz} as the base model for fine-tuning, which is trained on about 170,000 hours of single-sentence speech data.
2) \textbf{Second-Stage}: The model is trained with dialogue data to infer appropriate empathetic captions based on the dialogue context and to continue predicting the corresponding semantic codes.
Additionally, the Synthesizer can be trained separately in a single-sentence mode using empathetic captions and semantic codes, thereby enhancing the naturalness and robustness of the synthetic speech.
% The datasets used across the Multi Training will be described in the subsequent ``Datasets.''

% ///////////////////////////////////

\section{CSS-EmCap Pipeline}
In this section, we provide a detailed description of CSS-EmCap, which includes two components: 1) \textit{Multi-level Attribute Extraction.} 2) \textit{Empathetic Captions Generation}. Through this LLM-driven automatic dialog-aware pipeline, empathetic captions can be annotated for any CSS datasets.
% The following sections detail each step of the pipeline, with corresponding data flow diagrams available in the Appendix.

\subsection{Multi-level Attribute Extraction}
To improve the stability of LLMs in generating emotion- and style-related descriptions, we pre-extract two types of key expressive attributes (style factors and emotion) from conversational speech before generation. First, we use speech analysis tools to extract sentence-level style factors (including gender, pitch, energy, and tempo). Then we categorize the speech into different classes based on factor values and unified thresholds. Next, we use multimodal information such as speech, text, and speaker data to prompt LLM \footnote{https://deepmind.google/technologies/gemini/pro/ \label{gemini}} to accurately distinguish the emotional category of each sentence within the dialogue context.

\subsection{Empathetic Captions Generation}
After extracting style factors at the sentence level and emotions at the dialog level, we employ Gemini \footref{gemini} to generate diverse natural language descriptions for each speech. Unlike previous methods that rely solely on large language models to combine expressive attributes, we leverage Gemini’s speech understanding capabilities to create empathetic captions by integrating the original speech. The prompting process is divided into two main steps as follows: 1) \textbf{Step-1}: we generate basic descriptions based on the dialogue context and the two levels of extracted expressive attributes. 2) \textbf{Step-2}: we apply rules such as synonym replacement and varying emotional intensity descriptions to prompt Gemini to expand and enrich the captions. Additionally, we add a verification process to ensure that the descriptions accurately reflect the speech's expressiveness. Consequently, CSS-EmCap produces empathetic captions that more accurately convey the emotions and expressive styles present in the dialogue.

% ////////////////////////////////
\section{Experiments and Results}
In this section, we introduce the NCSSD-EmCap dataset used in this work, followed by a discussion of \textit{Baselines} and \textit{Metrics}. We then provide a comprehensive experimental analysis, including \textit{CSS-EmCap Evaluation}, \textit{Chain-Talker Evaluation}, \textit{Ablation Results}, \textit{Visualization Results}, and \textit{Hyperparameter Selection}. Further details on the \textit{Experimental Setup} and \textit{Case Study} are available in the Appendix.

\subsection{Datasets}
% The training data used in the Chain-Talker training process includes two types: 
% 1) \textit{Single-sentence speech synthesis data}: Each item data consists of (text, speech). In this work, we use ``CosyVoice-300M-25Hz'' \footnote{https://www.modelscope.cn/models/iic/CosyVoice-300M-25Hz} as the base model for fine-tuning, which is trained on about 170,000 hours of single-sentence speech data, including LibriTTS \cite{Heiga2019LibriTTS} and LJSpeech \cite{ito2017ljspeech}, among others.
% 2) \textit{Conversational speech synthesis data with empathetic caption annotation}: 
We employ the open-source DailyTalk \cite{lee2023dailytalk}, MultiDialog \cite{se2024multidialog} and NCSSD \cite{liu2024generative} datasets to develop the NCSSD-EmCap dataset via the CSS-EmCap pipeline. For detailed statistical information on these datasets, please refer to the Appendix.

% //////////////////////////////////
% 基线模型
% //////////////////////////////////

\subsection{Baselines}
To validate the effectiveness of the CSS-EmCap and the capabilities of Chain-Talker, we compare two categories of baseline models:

To validate the LLM-driven automatic dialog-aware empathetic captioning pipeline, we compare the following caption generation schemes:
1) \textbf{\textit{w/o SF}}: Direct use of the LLM \footref{gemini} to extract speech expressive attributes. 
2) \textbf{\textit{w/o SL-SF}}: Removal of sentence-level style factors, followed by caption generation using the LLM. 
3) \textbf{\textit{w/o DL-SF}}: Removal of dialog-level emotion, then using the LLM for caption generation. 
4) \textbf{\textit{Qwen2-Audio}}: LLM with speech style capture capabilities \cite{chu2024qwen2-audio}.
5) \textbf{\textit{SECap}}: LLM with speech emotion captions capture capabilities \cite{wu2024secap}.

% \textbf{Category $\text{II}$: Effectiveness of Chain-Talker in Dialogue Scenarios.} 
% To validate the Chain-Talker, we compare it with the following advanced CSS systems.
% 1) \textit{GRU-based Context Modeling}: we call ``CCATTS'' here \cite{guo2021conversational}, 
% 2) \textit{Multi-Scale Context Modeling}: ``M$^2$-CTTS'' \cite{xue2023M2ctts},
% 3) \textit{Heterogeneous Graph-based Context Modeling}: ``ECSS'' \cite{Liu2024ECSS}, 
% 4) \textit{GPT-based Context Modeling}: ``GPT-Talker'' \cite{liu2024generative}.
% 5) \textit{Label-based Modeling}: We introduce ``Chain-Talker$_s$'', a variant of Chain-Talker that replaces empathetic captions with style factor labels, and ``Chain-Talker$_e$'', which replaces empathetic captions with emotion labels.
% 6) \textit{No Context Modeling}: ``w/o context'', pure CosyVoice \cite{du2024cosyvoice}, 
% 7) \textit{No Emotion Understanding}: ``w/o captions'', which directly predicts speech tokens using the dialogue context and synthesizes the speech. 

We evaluate the effectiveness of Chain-Talker, trained based on NCSSD-EmCap, within dialogue scenarios by comparing it against state-of-the-art CSS systems:
1) \textbf{\textit{CCATTS}} \cite{guo2021conversational}, 
2) \textbf{\textit{M$^2$-CTTS}} \cite{xue2023M2ctts}, 
3) \textbf{\textit{ECSS}} \cite{Liu2024ECSS}, 
4) \textbf{\textit{GPT-Talker}} \cite{liu2024generative}, 
5) \textbf{\textit{GPT-Talker$_c$}} (GPT-Talker adds Emotion Understanding), 
6) \textbf{\textit{Chain-Talker$_e$}} (Using emotion labels to replace empathetic captions), 
7) \textbf{\textit{Chain-Talker$_s$}} (Using style labels to replace empathetic captions). 
Additionally, we also assess the importance of various modules and loss functions in Chain-Talker: 
8) \textbf{\textit{w/o context}}: A Chain-Talker variant without dialog history $\mathcal{H}$.
9) \textbf{\textit{w/o captions}}: A Chain-Talker variant without emotion understanding, only semantic understanding.
10) \textbf{\textit{w/o $\mathcal{L}^{caption}$}}: Removing the loss function about emotion understanding.
11) \textbf{\textit{w/o First-Stage}}: Training Chain-Talker directly on the NCSSD-EmCap dataset.

For details of the baseline models, please refer to the Appendix.

% ////////////////////////
% 评估指标
% ///////////////////////

\subsection{Metrics}

% \subsubsection{
\textbf{Objective Evaluation Metrics:}
1) \textit{Semantic Similarity (SIM$_*$):} We utilize RoBERTa \cite{liu2019roberta} and mGTE \cite{zhang2024mgte} to encode captions generated with different methods and descriptions composed of all Ground Truth style factors (e.g., ``gender is female, pitch is high..."). The semantic similarity is then calculated using cosine similarity. Higher values mean that the captions more accurately reflect the real style.
2) \textit{Caption Diversity (DIS-1/DIS-2)}: We use distinct-1/-2 \cite{li2015diversity} to evaluate the diversity of generated captions.
3) \textit{Emotion Accuracy (ACC$_m$):} We calculate the emotion accuracy of synthesized speech using Gemini.
% including gender (ACC$_g$), energy (ACC$_e$), pitch (ACC$_p$), tempo (ACC$_t$), and
4) \textit{Speaker Similarity (SSIM):}
Following \citet{jiang2023mega}, we use embeddings extracted from a fine-tuned WavLM \footnote{https://huggingface.co/microsoft/wavlm-base-plus-sv} model to assess speaker similarity of synthesized speech.
5) \textit{Dynamic Time Warping Distance (DDTW):}
We use the method from \citet{muller2007information} to measure expressiveness in speech by calculating the average Dynamic Time Warping distance of pitch distributions between real and synthesized speech, where lower values suggest higher similarity to Ground Truth.

\textbf{Subjective Evaluation Metrics:}
1) \textit{Dialog-level Mean Opinion Score for Naturalness (DMOS-N):} Participants are asked to judge the naturalness and quality of synthesized speech based on the dialogue context. 
2) \textit{Dialog-level Mean Opinion Score for Expressiveness (DMOS-E):} Participants are asked to evaluate whether the emotion and style of the synthesized speech match the current dialogue context.
3) \textit{Dialog-level Mean Opinion Score for Captions (DMOS-C):} Participants are asked to evaluate whether the generated empathetic captions match the given speech and dialogue context.

\begin{table}[t]
% \caption{\label{tab:exp-1}\textcolor{black}{Subjective (with 95\% confidence interval) and objective experimental results on the quality and diversity of empathetic captions.}} 
\caption{\label{tab:exp-1} Subjective (with 95\% confidence interval) and objective experimental results on the quality and diversity of empathetic captions. ``\textit{-w/o}'' indicates the removal of sub-steps within CSS-EmCap, where: ``\textit{SF}'' represents multi-level attribute extraction, ``\textit{SL-SF}'' denotes sentence-level style attribute extraction, and ``\textit{DL-SF}'' signifies dialogue-level emotion extraction.}

\centering
\resizebox{\linewidth}{!}{
\begin{tabular}{@{}
>{\columncolor[HTML]{FFFFFF}}l |  % 仍然保持第一列为左对齐
>{\columncolor[HTML]{FFFFFF}}c
>{\columncolor[HTML]{FFFFFF}}c
>{\columncolor[HTML]{FFFFFF}}c |
>{\columncolor[HTML]{FFFFFF}}c 
>{\columncolor[HTML]{FFFFFF}}c @{}}
\toprule
\multicolumn{1}{c|}{\cellcolor[HTML]{FFFFFF}\textbf{Methods}} &  % 这里使用 \multicolumn 来居中“Methods”
\multicolumn{1}{c}{\cellcolor[HTML]{FFFFFF}\textbf{DMOS-C $(\uparrow)$}} &
\multicolumn{1}{c}{\cellcolor[HTML]{FFFFFF}\textbf{SIM$_R$ $(\uparrow)$}} &
\multicolumn{1}{c|}{\cellcolor[HTML]{FFFFFF}\textbf{SIM$_G$ $(\uparrow)$}} &
\multicolumn{1}{c}{\cellcolor[HTML]{FFFFFF}\textbf{DIS-1 $(\uparrow)$}} &
\multicolumn{1}{c}{\cellcolor[HTML]{FFFFFF}\textbf{DIS-2 $(\uparrow)$}}
\\ \midrule
Ground Truth & \colorbox[RGB]{220,211,226}{4.327 $\pm$ 0.013} & -  & -  & -  & -   \\
Qwen2-Audio & 4.212 $\pm$ 0.018 & 0.431  & 0.534 & \underline{0.086}  & 0.174 \\
SECap & \underline{4.268 $\pm$ 0.022} & \underline{0.475}  & \underline{0.617} & 0.081  & \underline{0.186} \\
CSS-EmCap & 
    \colorbox[RGB]{221,232,250}{\textbf{4.462 $\pm$ 0.019}}& 
    \colorbox[RGB]{221,232,250}{\textbf{0.568}} &
    \colorbox[RGB]{221,232,250}{\textbf{0.694}} &
    \colorbox[RGB]{221,232,250}{\textbf{0.106}} & \colorbox[RGB]{221,232,250}{\textbf{0.296}} \\
~~~ -w/o SF & 3.819 $\pm$ 0.023 & 0.425  & 0.584  & 0.078  & 0.157   \\
~~~ -w/o SL-SF & 4.021 $\pm$ 0.017 & 0.335  & 0.384 & 0.024 & 0.049 \\
~~~ -w/o DL-SF & 4.113 $\pm$ 0.031 & 0.394  & 0.541 & 0.051 & 0.135 \\
\bottomrule
\end{tabular}
}
\end{table}

% ==================数据集的实验===================
\subsection{CSS-EmCap Evaluation}
We conduct a comprehensive analysis and evaluation of the annotated NCSSD-EmCap dataset, including the quality of the captions and the diversity of the generated descriptive styles.

\begin{table*}[t]
% \caption{\label{tab:exp-3}\textcolor{black}{Subjective (with 95\% confidence interval) and objective results with different CSS systems.}} 
\caption{\label{tab:exp-2} Subjective (with 95\% confidence interval) and objective results using different dialogue speech synthesis models. ``\textit{-w/o}'' indicates the removal of sub-modules within Chain-Talker, where: ``\textit{context}'' refers to dialogue context, ``\textit{captions}'' denotes empathetic captions, ``\textit{$\mathcal{L}^{caption}$}'' signifies the loss for the captions, and ``\textit{First-Stage}'' refers to the pre-training process using large-scale single-sentence data.}
\centering
\resizebox{0.75\linewidth}{!}{
\begin{tabular}{@{}l|ccccccccc@{}}
\toprule
\multicolumn{1}{c|}{\textbf{Methods}} &
  \textbf{DMOS-N $(\uparrow)$} &
  \textbf{DMOS-E $(\uparrow)$} &
  \textbf{ACC$_m$ $(\uparrow)$} &
  \textbf{DDTW $(\downarrow)$} &
  \textbf{SSIM $(\uparrow)$} \\ \midrule
Ground Truth & \colorbox[RGB]{220,211,226}{4.467 $\pm$ 0.020} & \colorbox[RGB]{220,211,226}{4.571 $\pm$ 0.015} & - & - & - \\
CCATTS  & 3.423 $\pm$ 0.033 & 3.469 $\pm$ 0.024 & 0.462 & 67.851 & 0.765 \\
M$^2$-CTTS & 3.461 $\pm$ 0.018 & 3.479 $\pm$ 0.021 & 0.471 & 66.184 & 0.769 \\
ECSS    & 3.655 $\pm$ 0.035 & 3.672 $\pm$ 0.029 & 0.495 & 59.749  & 0.785 \\
GPT-Talker  & 3.962 $\pm$ 0.011 & 3.913 $\pm$ 0.028 & 0.562 & 44.625 & 0.814 \\
GPT-Talker$_c$  & \underline{4.045 $\pm$ 0.021} & 4.102 $\pm$ 0.013 & 0.589 & \underline{40.374} & 0.829 \\
Chain-Talker$_{s}$  & 4.036 $\pm$ 0.027 & 4.015 $\pm$ 0.026 & 0.578 & 42.876 & \underline{0.851} \\
Chain-Talker$_{e}$  & 4.022 $\pm$ 0.019 & \underline{4.127 $\pm$ 0.021} & \underline{0.601} & 40.763 & 0.849 \\
\textbf{Chain-Talker} &
  \colorbox[RGB]{221,232,250}{\textbf{4.147 $\pm$ 0.024}} &
  \colorbox[RGB]{221,232,250}{\textbf{4.239  $\pm$ 0.011}} &
  \colorbox[RGB]{221,232,250}{\textbf{0.612}} &
  \colorbox[RGB]{221,232,250}{\textbf{38.784}} &
  \colorbox[RGB]{221,232,250}{\textbf{0.862}} \\ \midrule
~~~-w/o context & 3.982 $\pm$ 0.038 & 3.984 $\pm$ 0.014 & 
0.564 & 43.589 & 0.847 \\
~~~-w/o captions  &  4.037 $\pm$ 0.021 & 4.084 $\pm$ 0.032 & 0.571 & 43.479 & 0.836 \\
~~~-w/o $\mathcal{L}^{caption}$ & 3.947 $\pm$ 0.015 & 3.956 $\pm$ 0.025 & 0.568 & 45.764 & 0.829 \\ 
~~~-w/o First-Stage & 3.756 $\pm$ 0.024 & 3.789 $\pm$ 0.018 & 0.517 &52.640 & 0.793 \\
\bottomrule
\end{tabular}
}
\end{table*}

\textbf{Quality Evaluation:}
To evaluate whether captions generated by different methods accurately capture the emotions and styles of conversational speech, we randomly select 40 dialogue sets (comprising a total of 240 utterances) from the NCSSD-EmCap dataset. Subsequently, we employ the Category $\text{I}$ baseline models described earlier to generate captions for the 240 utterances. Finally, we compare these captions with the Ground Truth to compute \textit{SIM$_*$} and conduct \textit{DMOS-C} evaluations. Particularly, this \textit{DMOS-C} evaluation involves 30 university students who are proficient in English as a second language and have strong dialogue and reading skills.
As shown in the second to fourth columns of Table \ref{tab:exp-1}, by comparing \textit{DMOS-C} and \textit{SIM$_*$}, it is evident that our data annotation scheme exhibits clear advantages over other methods. Furthermore, the observation that \textit{DMOS-C} values surpass those of the Ground Truth demonstrates that empathetic captions described in natural language are superior to style and emotion labels.    

\textbf{Diversity Evaluation:}
To evaluate the diversity of annotated empathetic captions, within the NCSSD-EmCap dataset, we identify 10 different style combinations based on various style factors and emotions. For each combination, we select 50 corresponding captions. We compute the \textit{distinct-1} and \textit{distinct-2} values for the 50 captions of each style, and the average values across all 10 styles are calculated as the experimental results. 
As shown in the fifth and sixth columns of Table \ref{tab:exp-1}, our scheme outperforms others with scores of 0.106 and 0.296, respectively. This demonstrates that our designed pipeline, leveraging LLMs, can generate diverse and appropriate natural language descriptions.

\begin{figure*}[th]
\centering
\centerline{
\includegraphics[width=0.9\linewidth]{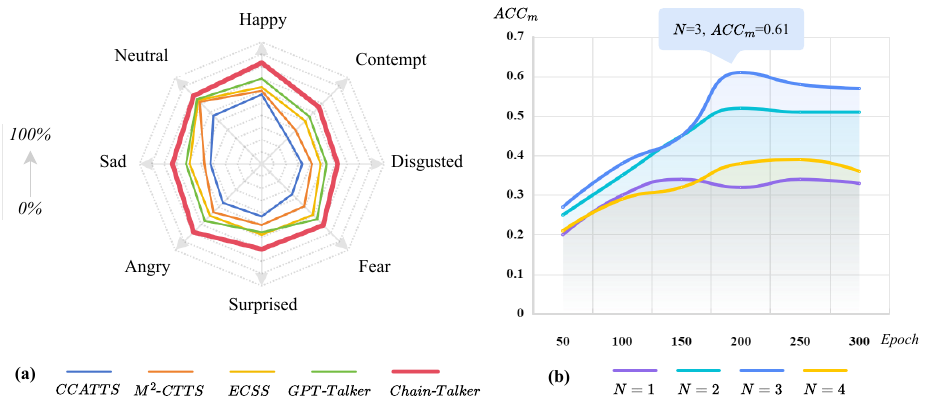}
}
\caption{(a) The ability of Chain-Talker and various baseline models to synthesize target speech with different emotional categories based on dialogue context. (b) Experimental results on the selection of the hyperparameter $N$ for dialogue turns.}
\label{fig:emotion}
\end{figure*}

% ===================模型的实验====================
\subsection{Chain-Talker Evaluation}
We compare Chain-Talker with seven advanced CSS models using the NCSSD-EmCap dataset. For subjective evaluation, 30 university students rate 50 randomly selected synthesized sentences from the test set. The ratings are given in a quiet environment based on the context and using the criteria \textit{DMOS-N} and \textit{DMOS-E}. For objective evaluation, we measure the \textit{ACC$_m$}, \textit{DDTW}, and \textit{SSIM} for each model's synthesized speech. The results are presented in rows two through ten of Table \ref{tab:exp-2}.
The results show that Chain-Talker achieves significantly lower \textit{DDTW} scores compared to other CSS models, indicating that its combination of GPT-based context modeling and CFM-based acoustic modeling excels in capturing pitch-related expressiveness. Additionally, the \textit{SSIM} scores suggest that Chain-Talker's synthesized speech closely resembles the Ground Truth, effectively preserving the agent's timbre. The \textit{ACC$_m$} metric demonstrates that Chain-Talker is able to empathize by understanding and rendering generated speech emotions that are more appropriate for the entire conversation. 
In subjective evaluations, Chain-Talker ranks higher than its closest competitor by 0.102 in naturalness MOS and leads by at least 0.112 in expressiveness MOS. Overall, these results confirm Chain-Talker's ability to both understand dialogue context and produce empathetic, context-aligned speech.

Furthermore, GPT-Talker$_c$’s results show that adding emotional understanding and expression effectively enhances the model’s empathy. However, its speech encoding, derived from HuBERT, includes some acoustic information alongside semantics, which impacts emotional and semantic comprehension. This further highlights the importance of using speech encoding with purely semantic information in our work.

% ====================其他实验====================
\subsection{Ablation Results}
To verify the effectiveness of our model components, we conduct ablation experiments by removing specific modules and training methods, with results shown in rows eleven to fourteen of Table \ref{tab:exp-2}.
The ``w/o context'' condition, which excludes the dialogue history, shows significant difficulty in context-aware speech synthesis compared to Chain-Talker, indicating the importance of context modeling.
``w/o captions'' further shows that our Emotion Understanding and Empathetic Rendering are more effective than merely predicting and decoding speech token sequences. 
Under the ``w/o \( \mathcal{L}^{caption} \)'' condition, omitting the caption loss leads to a performance drop (\textit{DMOS-N} decreases by 0.2 and \textit{DMOS-E} by 0.283), highlighting its impact on inference stability and empathetic captions' accuracy. Comparing Chain-Talker with the results of ``w/o First-Stage'' indicates that pretraining the model on large-scale single-sentence data and then fine-tuning it on small-scale dialogue data can achieve better performance.

\subsection{Visualization Results}
To clearly demonstrate Chain-Talker’s ability to comprehend and convey emotions in dialogue-based speech synthesis, we select 240 target sentences from various emotional categories. Based on the dialogue context, we identify and calculate the emotional categories of the synthesized speech for different models and present their accuracy rates in Fig.~\ref{fig:emotion}. In this figure, the arrows indicate each model’s strength in understanding and rendering the corresponding emotion. The results show that Chain-Talker generally outperforms other baseline models in emotional expression, further confirming the superiority of chain modeling.

\subsection{Hyperparameter Selection}
To clearly demonstrate the impact of dialogue context length 
$N$ on the model's empathy, we randomly select 100 dialogue pairs from the NCSSD-EmCap dataset. We assess the emotion accuracy of the synthesized speech produced by the model across different training epochs, setting $N$ to $1 \rightarrow 3$ during training and $1 \rightarrow 4$ during inference. Fig. \ref{fig:emotion} illustrates that Chain-Talker's ability to understand and express emotions improves significantly with more training epochs, achieving peak performance around 200 epochs. The best results occur when $N$=3 at approximately 200 epochs. Although performance slightly declines at $N$=4, comparisons with $N$=1 demonstrate that the model can still manage dialogue lengths not encountered during training. Consequently, it is reasonable to hypothesize that increasing the number of dialogue turns during training could potentially enhance the Chain-Talker’s performance.

\section{Conclusion}
In this work, we introduce a novel chain modeling-based CSS model named Chain-Talker, designed for spoken interaction in user-agent communications. This model comprises three primary processes: ``Emotion Understanding'', ``Semantic Understanding'', and ``Empathetic Rendering''. This step-by-step modeling of the conversational process significantly reduces the difficulty of responding to empathetic speech.
Additionally, we also develop an LLM-driven automated dialog-aware empathetic caption generation pipeline called CSS-EmCap. This pipeline has been utilized to annotate three open-source CSS datasets, DailyTalk, NCSSD, and MultiDialog. These datasets provide strong support for the training of Chain-Talker. The annotation pipeline will be made openly available, fostering community development.

% /////////////////////////////////////////////

\section*{Limitations}
\textbf{Inference Latency:} We conduct inference tests using an NVIDIA GeForce RTX 4080 GPU with 32 GB of VRAM and a 12th Gen Intel® Core™ i7-12700K CPU with 32 GB of system RAM. The average duration of empathetic speech responses generated by Chain-Talker is 2.5 seconds, which we consider acceptable. However, there remains a gap compared to real-time interactions. In future work, we continue to explore faster dialogue modeling methods that incorporate empathetic capabilities, such as the integration of streaming inference.

\textbf{Robustness:} We employ ``CosyVoice-300M-25Hz'' as the foundational model, which is pretrained on approximately 170,000 hours of speech data and demonstrates exceptionally high naturalness in synthesized speech. However, the conversational data used for fine-tuning comprises only 384 hours, predominantly featuring young speakers. As a result, Chain-Talker may not accurately capture the conversational styles of children and the elderly. In the future, we plan to construct larger-scale speech dialogue datasets to further enhance the model's robustness.

% \textcolor{red}{
% Since December 2023, a "Limitations" section has been required for all papers submitted to ACL Rolling Review (ARR). This section should be placed at the end of the paper, before the references. The "Limitations" section (along with, optionally, a section for ethical considerations) may be up to one page and will not count toward the final page limit. Note that these files may be used by venues that do not rely on ARR so it is recommended to verify the requirement of a "Limitations" section and other criteria with the venue in question.}

\section*{Ethics Statement}
Safety Risks: Chain-Talker possesses zero-shot speech synthesis capabilities, facilitating the creation of personalized conversational speech. In most cases, individuals are likely to utilize this technology to enhance movie dubbing, podcasts, and other services. However, it may also present potential risks for model misuse, such as spoofing voice. To address this, we plan to incorporate restrictions into the open-source license of the Chain-Talker project to prevent the misuse of the model.

\section*{Acknowledgment}
The research by Rui Liu was funded by the Young Scientists Fund (No.~62206136), the General Program (No.~62476146) of the National Natural Science Foundation of China, and the Young Elite Scientists Sponsorship Program by CAST (2024QNRC001).
The research by Yifan Hu was funded by the Research and
Innovation Projects for Graduate Students in Inner Mongolia Autonomous Region.
The work by Haizhou Li was supported by the Shenzhen Science and Technology Program (Shenzhen Key Laboratory, Grant No.~ZDSYS20230626091302006), the Shenzhen Science and Technology Research Fund (Fundamental Research Key Project, Grant No.~JCYJ20220818103001002), and the Program for Guangdong Introducing Innovative and Enterpreneurial Teams, Grant No.~2023ZT10X044.

% Bibliography entries for the entire Anthology, followed by custom entries
%\bibliography{anthology,custom}
% Custom bibliography entries only
\bibliography{main}

\clearpage

\appendix

\section*{Technical Appendix}
\label{sec:appendix}
In this technical appendix, we will supplement the description of the proposed CSS model: Chain-Talker, the implementation details of the LLM-driven automatic dialog-aware
empathetic caption generation pipeline: CSS-EmCap, and additional experimental results.

% 模型的更多细节
% \section{More Details of Chain-Talker}

% pipeline的更多细节
\section{More Details of CSS-EmCap}
\subsection{Data Flow Diagram of CSS-EmCap}
\begin{figure*}[t]
\centering
\centerline{
\includegraphics[width=1\linewidth]{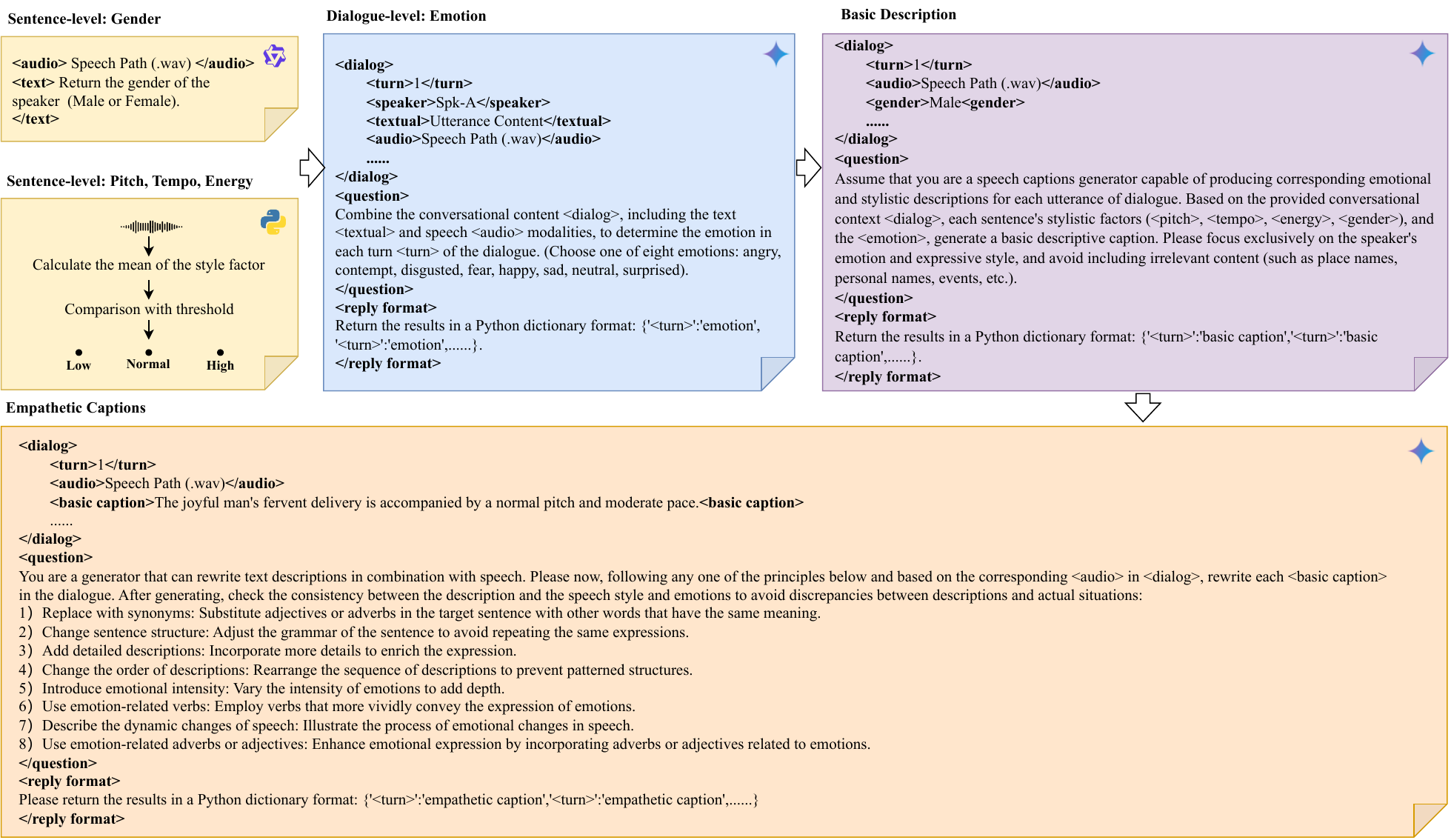}
}
\caption{The overall process of CSS-EmCap, It includes extracting Sentence-level style factors and Dialog-level emoton, as well as prompting LLM to generate Basic Descriptions and final Empathetic Captions.}
\label{fig:emcap}
\end{figure*}

As shown in Fig.~\ref{fig:emcap}, the empathetic caption annotation process in CSS-EmCap comprises several steps. First, we extract sentence-level style factors—such as gender, pitch, energy, and tempo—along with dialog-level emotions: 
\begin{itemize}
    \item \textbf{Sentence-level Style Factors Extraction},we employ the Qwen2-Audio large-scale model \footnote{https://huggingface.co/Qwen/Qwen2-Audio-7B} for speaker gender recognition, which achieves superior accuracy by providing the speech path in the prompt and having the model return the corresponding result (male or female). Additionally, Librosa \footnote{https://librosa.org/} is used to analyze energy levels, and the World Vocoder \cite{morise2016world} extracts pitch information. Furthermore, MFA \footnote{https://montreal-forced-aligner.readthedocs.io/en/latest/} aligns text and speech data to obtain duration information, which is then averaged. To ensure that datasets annotated using this pipeline adhere to the same hierarchical classification standards, a set of uniform thresholds \footnote{The three sets of thresholds are: pitch [136.577, 196.098], tempo [0.252, 0.386], and energy [0.033, 0.0505]. Within each set, the first value distinguishes between low and normal, and the second between normal and high.} is established after calculating the factor values for all speech data. Pitch, energy, and tempo are categorized into three levels: low, normal, and high.
    \item \textbf{Dialog-level Emotion Extraction}, in order to identify the emotional category of each speech in the dialogue, we carefully design prompts to enable the state-of-the-art Gemini 1.5 pro model \footref{gemini}, which understands and analyzes both text and speech modalities, to return results accurately. The model receives the complete dialogue content, encapsulated within $\langle dialog \rangle ...\langle /dialog \rangle$, including dialogue turns, speaker, textual content, and corresponding speech paths. It then identifies the emotional category of each speech, leveraging the context provided by the entire dialogue. Compared to pure single-sentence emotion recognition models, this context-based approach improves the accuracy of emotion recognition.
\end{itemize}

Next, we employ a LLM \footref{gemini} to generate basic descriptions by integrating these extracted expressive attributes with the dialogue's speech. Subsequently, we utilize the same LLM to expand these descriptions using one of eight predefined rules. Simultaneously, it performs consistency checks to ensure that the descriptions accurately reflect the original speech, resulting in the final empathetic captions.

\subsection{Statistical Information of NCSSD-EmCap}
As shown in Table \ref{tab:statistic}, the overall statistics for the NCSSD-EmCap dataset are presented. It consists of 384 hours of natural spoken dialogue, comprising 18,580 dialogs and 245,984 dialogue utterances. Each utterance includes speaker information, text, empathetic captions, and corresponding speech. DailyTalk, NCSSD, and MultiDialog offer a wide range of emotional categories along with varying levels of pitch, energy, and tempo, providing strong support for training highly expressive CSS models.

% //////////////////////////////////////
\begin{table}[t]
\caption{\label{tab:statistic}\textcolor{black}{Statistical results for the NCSSD-EmCap dataset, includes DailyTalk, NCSSD, and MultiDialog subsets.}} 
\centering
\resizebox{1\linewidth}{!}{
\begin{tabular}{cccccc}
\hline
\multicolumn{6}{c}{\textbf{NCSSD-EmCap}}    \\ \hline
\multicolumn{1}{c|}{\multirow{2}{*}{\textbf{Factors}}} &
  \multicolumn{1}{c|}{\multirow{2}{*}{\textbf{Items}}} &
  \multicolumn{3}{c|}{\textbf{Sub-datasets}} & \multirow{2}{*}{
  \textbf{Total}} \\ \cline{3-5}
\multicolumn{1}{c|}{} &
  \multicolumn{1}{c|}{} &
  \textbf{DailyTalk} &
  \textbf{NCSSD} &
  \multicolumn{1}{c|}{\textbf{MultiDialog}} &
   \\ \hline
\multicolumn{1}{c|}{\multirow{2}{*}{Gender}}  & \multicolumn{1}{c|}{Male}      & 11,866 & 33,672 & \multicolumn{1}{c|}{79,061} & 124,599 \\
\multicolumn{1}{c|}{}                         & \multicolumn{1}{c|}{Female}    & 11,906 & 38,964 & \multicolumn{1}{c|}{70,515} & 121,385 \\ \hline
\multicolumn{1}{c|}{\multirow{3}{*}{Pitch}}   & \multicolumn{1}{c|}{Low}       & 4,594  & 18,444 & \multicolumn{1}{c|}{35,787} & 58,825 \\
\multicolumn{1}{c|}{}                         & \multicolumn{1}{c|}{Normal}    & 8,143  & 22,495 & \multicolumn{1}{c|}{41,973} & 72,611 \\
\multicolumn{1}{c|}{}                         & \multicolumn{1}{c|}{High}      & 11,035 & 31,697 & \multicolumn{1}{c|}{71,816} & 114,548 \\ \hline
\multicolumn{1}{c|}{\multirow{3}{*}{Energy}}  & \multicolumn{1}{c|}{Low}       & 2      & 72,636 & \multicolumn{1}{c|}{18,546} & 91,184 \\
\multicolumn{1}{c|}{}                         & \multicolumn{1}{c|}{Normal}    & 52     & 0      & \multicolumn{1}{c|}{33,650} & 33,702 \\
\multicolumn{1}{c|}{}                         & \multicolumn{1}{c|}{High}      & 23,718 & 0      & \multicolumn{1}{c|}{97,380} & 121,098 \\ \hline
\multicolumn{1}{c|}{\multirow{3}{*}{Tempo}}   & \multicolumn{1}{c|}{Low}       & 4,048  & 12,491 & \multicolumn{1}{c|}{2,780} & 19,319 \\
\multicolumn{1}{c|}{}                         & \multicolumn{1}{c|}{Normal}    & 15,980 & 25,262 & \multicolumn{1}{c|}{76,282} & 117,524 \\
\multicolumn{1}{c|}{}                         & \multicolumn{1}{c|}{High}      & 3,744  & 34,883 & \multicolumn{1}{c|}{70,514} & 109,141 \\ \hline
\multicolumn{1}{c|}{\multirow{8}{*}{Emotion}} & \multicolumn{1}{c|}{Angry}     & 378    & 8,048  & \multicolumn{1}{c|}{716} & 9,142 \\
\multicolumn{1}{c|}{}                         & \multicolumn{1}{c|}{Contempt}  & 24     & 390    & \multicolumn{1}{c|}{0} & 414 \\
\multicolumn{1}{c|}{}                         & \multicolumn{1}{c|}{Disgusted} & 57     & 277    & \multicolumn{1}{c|}{1,152} & 1,486 \\
\multicolumn{1}{c|}{}                         & \multicolumn{1}{c|}{Fear}      & 86     & 781    & \multicolumn{1}{c|}{760} & 1,627 \\
\multicolumn{1}{c|}{}                         & \multicolumn{1}{c|}{Happy}     & 2,613  & 6,026  & \multicolumn{1}{c|}{23,393} & 32,032 \\
\multicolumn{1}{c|}{}                         & \multicolumn{1}{c|}{Sad}       & 848    & 7,007  & \multicolumn{1}{c|}{1,975} & 9,830 \\
\multicolumn{1}{c|}{}                         & \multicolumn{1}{c|}{Neutral}   & 19,240 & 47,281 & \multicolumn{1}{c|}{97,708} & 164,229 \\
\multicolumn{1}{c|}{}                         & \multicolumn{1}{c|}{Surprised} & 526    & 2,826  & \multicolumn{1}{c|}{23,872} & 27,224 \\ \hline
\multicolumn{2}{c|}{\textbf{Hours}}                                            & 20     & 92     & \multicolumn{1}{c|}{272} & 384 \\ \hline
\multicolumn{2}{c|}{\textbf{Dialogs}}                                          & 2,541  & 8,229  & \multicolumn{1}{c|}{7,810} & 18,580 \\ \hline
\multicolumn{2}{c|}{\textbf{Utterances}}                                       & 23,772 & 72,636 & \multicolumn{1}{c|}{149,576} & 245,984 \\ \hline
\end{tabular}
 }
\end{table}

% //////////////////////////////////

% 实验的更多细节
\section{More Details of Experiments and Results}
\subsection{Baseline Models}
In this part, we will detail the baseline models compared in this work, still organized into two categories:

\textbf{Category $\text{I}$: Evaluation of the LLM-driven Automatic Dialog-aware Empathetic Captioning Pipeline CSS-EmCap.} 
\begin{itemize}
    \item \textit{\textbf{w/o SF}}: We directly use the LLM \footref{gemini} to generate captions from given speech without extracting style factors and emotion.
    \item \textit{\textbf{w/o SL-SF}}: We remove the sentence-related style factors, and then generate captions using only emotional labels with the LLM.
    \item \textit{\textbf{w/o DL-SF}}: We remove the extraction process of Dialog-level emotion, and simply use gender, pitch, tempo, and energy as prompts for the LLM to generate style captions for speech.
    \item \textit{\textbf{Qwen2-Audio}}: A versatile multi-task LLM that accepts both audio (including human speech, natural sounds, and music) and text inputs, and outputs text. This model has the ability to understand speech content and style \cite{chu2024qwen2-audio}.
    \item \textit{\textbf{SECap}}: A LLM trained on a 41-hour emotional dataset, capable of returning speech emotions described in natural language \cite{wu2024secap}.
\end{itemize}

\textbf{Category $\text{II}$: Effectiveness of Chain-Talker in Dialogue Scenarios.} 
\begin{itemize}
    \item \textit{\textbf{CCATTS}}: A context-aware CSS model, which employs a GRU-based network to model the sentence-level dependency among the dialogue context \cite{guo2021conversational}. 
    \item \textit{\textbf{M$^2$-CTTS}}: It extracts context-dependent information at multiple granularities—both word-level and sentence-level—from the speech and text within the dialogue context \cite{xue2023M2ctts}.
    \item \textit{\textbf{ECSS}}: It considers different utterances in the dialogue context and modal information as individual graph nodes and uses a heterogeneous graph neural network to model the dialogue context \cite{Liu2024ECSS}.
    \item \textit{\textbf{GPT-Talker}}: It takes the text and speech of the dialogue context as prompts and then utilizes a GPT-style autoregressive model to predict the discrete token sequence of the response speech, followed by synthesizing the final speech waveform using VITS \cite{liu2024generative, kim2021VITS}.
    \item \textit{\textbf{GPT-Talker$_c$}}: A variant of GPT-Talker adds empathetic captions during context modeling to help understand emotional changes in conversations.
    \item \textbf{\textit{Chain-Talker$_s$} and \textit{Chain-Talker$_e$}}: They are two variants of Chain-Talker, the former replacing the empathetic captions with style factor labels and the latter with emotion labels.
    \item \textit{\textbf{w/o context}}: A Chain-Talker variant without dialog history $\mathcal{H}$, it predicts caption and renders speech directly from the given target utterance.
    \item \textit{\textbf{w/o captions}}: A CosyVoice variant, similar to GPT-Talker, where the input part includes various modal information from the dialogue history in addition to the target sentence, but without empathetic captions.
\end{itemize}

\begin{table}[t]
\caption{\label{tab:setup}\textcolor{black}{Statistics of some model parameters.}} 
\centering
\resizebox{0.7\linewidth}{!}{
\begin{tabular}{c|lc}
\hline
\multirow{13}{*}{\textbf{EmGPT}} & \multicolumn{1}{l|}{speech sample\_rate} & 22050 \\ \cline{2-3} 
 & \multicolumn{1}{l|}{spk\_embed\_dim} & 192 \\ \cline{2-3} 
 & \multicolumn{1}{l|}{text\_token\_size} & 60515 \\ \cline{2-3} 
 & \multicolumn{1}{l|}{speech\_token\_size} & 4096 \\ \cline{2-3} 
 & \multicolumn{2}{c}{LLM} \\ \cline{2-3} 
 & \multicolumn{1}{l|}{llm\_input\_size} & 1024 \\ \cline{2-3} 
 & \multicolumn{1}{l|}{llm\_output\_size} & 1024 \\ \cline{2-3} 
 & \multicolumn{1}{l|}{num\_blocks} & 14 \\ \cline{2-3} 
 & \multicolumn{1}{l|}{dropout\_rate} & 0.1 \\ \cline{2-3} 
 & \multicolumn{2}{c}{Sampling} \\ \cline{2-3} 
 & \multicolumn{1}{l|}{top\_k} & 25 \\ \cline{2-3} 
 & \multicolumn{1}{l|}{win\_size} & 10 \\ 
 \cline{2-3} 
  & \multicolumn{1}{l|}{tau\_r} & 0.1 \\ 
 \hline
\multicolumn{1}{l|}{\multirow{11}{*}{\textbf{Synthesizer}}} & \multicolumn{2}{c}{OT-CFM} \\ \cline{2-3} 
\multicolumn{1}{l|}{} & \multicolumn{1}{l|}{input\_size} & 512 \\ \cline{2-3} 
\multicolumn{1}{l|}{} & \multicolumn{1}{l|}{output\_size} & 80 \\ \cline{2-3} 
\multicolumn{1}{l|}{} & \multicolumn{1}{l|}{output\_type} & mel \\ \cline{2-3} 
\multicolumn{1}{l|}{} & \multicolumn{1}{l|}{vocab\_size} & 4096 \\ \cline{2-3} 
\multicolumn{1}{l|}{} & \multicolumn{1}{l|}{input\_frame\_rate} & 25 \\ \cline{2-3} 
\multicolumn{1}{l|}{} & \multicolumn{2}{c}{HiFiGAN} \\ \cline{2-3} 
\multicolumn{1}{l|}{} & \multicolumn{1}{l|}{in\_channels} & 80 \\ \cline{2-3} 
\multicolumn{1}{l|}{} & \multicolumn{1}{l|}{base\_channels} & 512 \\ \cline{2-3} 
\multicolumn{1}{l|}{} & \multicolumn{1}{l|}{upsample\_rates} & {[}8, 8{]} \\ \cline{2-3} 
\multicolumn{1}{l|}{} & \multicolumn{1}{l|}{upsample\_kernel\_sizes} & {[}16, 16{]} \\ \hline
\end{tabular}
}
\end{table}

\subsection{Experimental Setup}
For some parameter settings of the two modules EmGPT and Synthesizer in the model Chain-Talker, please refer to Table \ref{tab:setup}. For additional details on the model configuration, please refer to our open-source repository on GitHub. Moreover, the Chain-Talker model was trained on four NVIDIA A800s. All datasets used are divided into training, validation, and test sets with a ratio of 8:1:1. During training, Chain-Talker's dialog turns are set to one to three. To ensure fairness in experimental results, during inference, the dialogue is also set to three turns for Chain-Talker and other CSS baseline models. For decoding strategy, EmGPT uses an auto-regressive decoding method, specifically employing the Top-K sampling strategy. 
% For model size, EmGPT is 1.41 GB and Synthesizer is 478.56 MB.

\begin{table}[t!]
\caption{\label{tab:exp-style}\textcolor{black}{Comparative results on emotion and style controllability.}} 
\centering
\resizebox{1\linewidth}{!}{
\begin{tabular}{cccccc}
\hline
\multicolumn{6}{c}{\textbf{Dataset: NCSSD-EmCap}}                                     \\ \hline
\multicolumn{1}{c|}{\textbf{Methods}} & \textbf{ACC$_g$ $(\uparrow)$} & \textbf{ACC$_e$ $(\uparrow)$} & \textbf{ACC$_p$ $(\uparrow)$} & \textbf{ACC$_t$ $(\uparrow)$} & \textbf{ACC$_m$ $(\uparrow)$} \\ \hline
\multicolumn{1}{c|}{PromptTTS}  & 0.825 & 0.728  & 0.826 & 0.739 & 0.487 \\
\multicolumn{1}{c|}{Salle} & \underline{0.841}           & \colorbox[RGB]{221,232,250}{\textbf{0.766}}  & \underline{0.852}  & \underline{0.742} & \underline{0.516} \\
\multicolumn{1}{c|}{Chain-Talker} & \colorbox[RGB]{221,232,250}{\textbf{0.854}}  & \underline{0.759} & \colorbox[RGB]{221,232,250}{\textbf{0.861}} & \colorbox[RGB]{221,232,250}{\textbf{0.747}}  & \colorbox[RGB]{221,232,250}{\textbf{0.623}} \\ 
\hline
\multicolumn{6}{c}{\textbf{Dataset: TextrolSpeech}}                             \\ \hline
\multicolumn{1}{c|}{\textbf{Methods}} & \textbf{ACC$_g$ $(\uparrow)$} & \textbf{ACC$_e$ $(\uparrow)$} & \textbf{ACC$_p$ $(\uparrow)$} & \textbf{ACC$_t$ $(\uparrow)$} & \textbf{ACC$_m$ $(\uparrow)$} \\ \hline
\multicolumn{1}{c|}{PromptTTS} & 0.834 & 0.746 & 0.835 & \underline{0.744} & 0.494 \\
\multicolumn{1}{c|}{Salle} & \underline{0.856} & \underline{0.761}  & \underline{0.836} & \colorbox[RGB]{221,232,250}{\textbf{0.751}} & \underline{0.506} \\
\multicolumn{1}{c|}{Chain-Talker} & \colorbox[RGB]{221,232,250}{\textbf{0.864}}  & \colorbox[RGB]{221,232,250}{\textbf{0.765}} & \colorbox[RGB]{221,232,250}{\textbf{0.857}}  & 0.743  & \colorbox[RGB]{221,232,250}{\textbf{0.598}}  \\ \hline
\end{tabular}
}
\end{table}

\begin{table}[t]
\caption{\label{tab:exp-case-1}\textcolor{black}{A sample set of empathetic captions generated by Chain-Talker at different epochs.}} 
\centering
\resizebox{\linewidth}{!}{
\begin{tabular}{c|c}
\hline
\textbf{Source} & \textbf{Emapthetic Caption} \\ \hline
\textbf{Ground Truth} & \begin{tabular}[c]{@{}c@{}}The speaker, a 
\redDashedBox{wrathful}
\colorbox[RGB]{217,231,214}{man}, \\ speaks with a \colorbox[RGB]{241,208,205}{lively tone} at a \colorbox[RGB]{223,213,230}{moderate pace}, \\ radiating \colorbox[RGB]{221,232,250}{high energy}.\end{tabular} \\ \hline
\textbf{Epoch 50} & In \colorbox[RGB]{217,231,214}{his} speech, the tone is high and \colorbox[RGB]{221,232,250}{energetic}. \\ \hline
\textbf{Epoch 160} & \begin{tabular}[c]{@{}c@{}}The \redDashedBox{wrathful} \colorbox[RGB]{217,231,214}{male} speaker addresses with lively speech, \\ \colorbox[RGB]{241,208,205}{normal pitch}.\end{tabular} \\ \hline
\textbf{Epoch 200} & \begin{tabular}[c]{@{}c@{}}\colorbox[RGB]{217,231,214}{His} voice, though at a \colorbox[RGB]{241,208,205}{\colorbox[RGB]{223,213,230}{normal} pitch} and \colorbox[RGB]{223,213,230}{speed}, \\  \redDashedBox{is filled with a palpable sense of fury}.\end{tabular} \\ \hline
\end{tabular}
}
\end{table}

\begin{figure*}[t]
\centering
\centerline{
\includegraphics[width=1\linewidth]{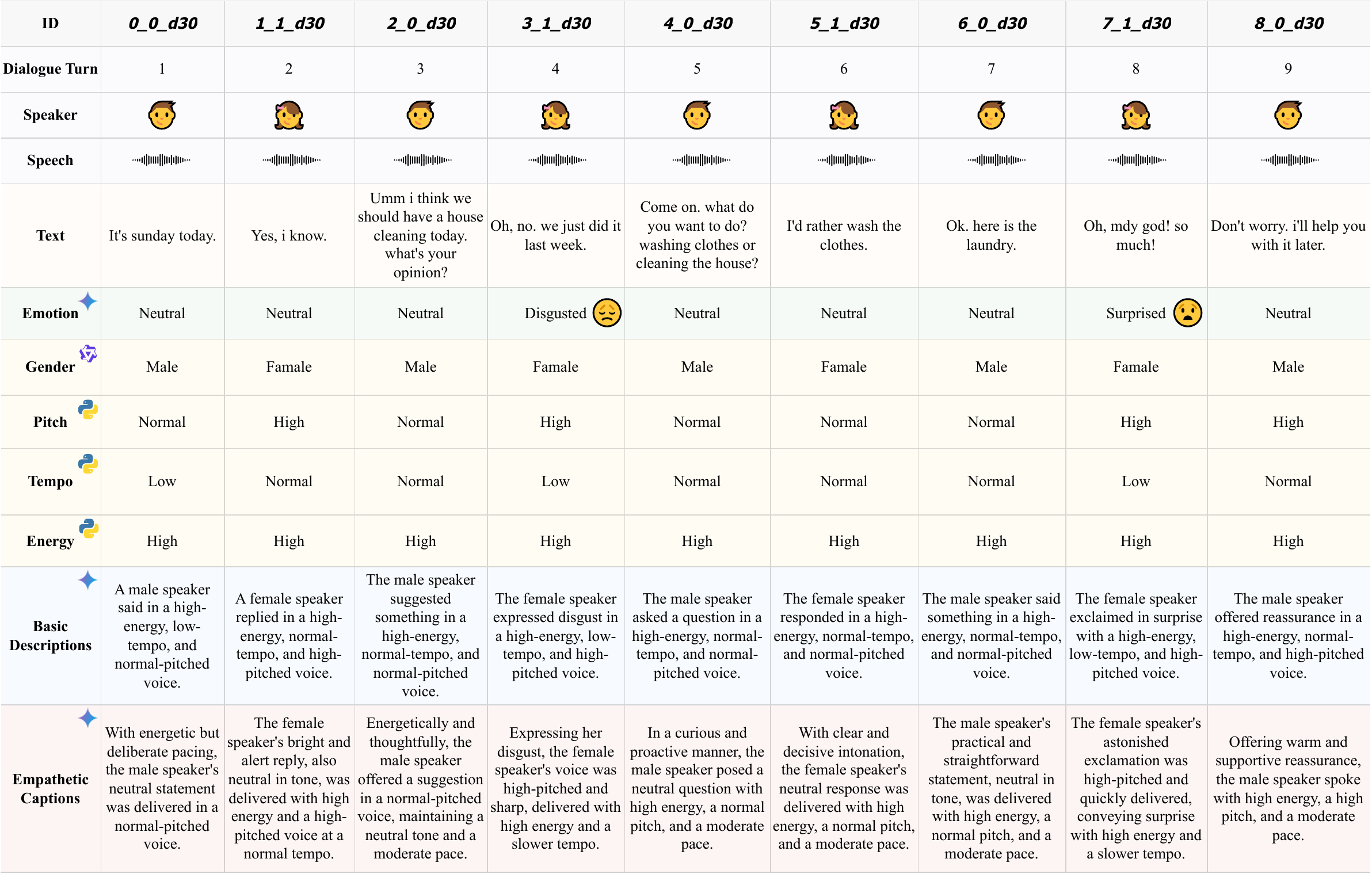}
}
\caption{A sample set of conversational data annotated using the CSS-EmCap pipeline.}
\label{fig:exp-case-2}
\end{figure*}

\subsection{Verifying Emotion and Style Controllability}
In Chain-Talker, the consistency between empathetic captions and dialogue speech in terms of emotion and style is crucial, as it directly affects the context modeling and rendering processes. Therefore, in this part, we aim to evaluate whether Chain-Talker can synthesize speech with the corresponding style and emotion directly from a given caption in a single-sentence mode. We conduct comparative experiments with the two most advanced natural language-controlled style TTS models, PromptTTS \cite{Guo2023prompttts} and Salle \cite{ji2024textrolspeech}. The evaluation metrics are consistent with the previously used $ACC_m$ (Emotion) and include $ACC_p$ (Pitch), $ACC_e$ (Energy), $ACC_t$ (Tempo), and $ACC_g$ (Gender). These metrics are used to determine the accuracy of the synthesized speech in conveying different expressive attributes.
We evaluate three models using the constructed NCSSD-EmCap dataset, please refer to Table~\ref{tab:exp-style}. Chain-Talker outperforms the other models in all metrics except energy. Additionally, we also evaluate them on the open-source style-controlled dataset TextrolSpeech \cite{ji2024textrolspeech}. The results demonstrate that Chain-Talker consistently achieves outstanding performance, particularly excelling in emotional expression by 0.092\% compared to the second-best model. All experimental results demonstrate that Chain-Talker has effective style control capabilities. In other words, as long as an appropriate empathetic caption is predicted, it can generate speech with the corresponding emotion and style in conversational settings. This also provides additional evidence supporting our previous experiments that evaluated Chain-Talker’s performance in CSS scenarios.

\subsection{Details in Subjective Evaluation}
In this study, we recruited 30 university students to participate in subjective evaluations, compensating them at a rate of \$15 per experiment. This remuneration is fair and reasonable locally. For \textit{DMOS-N}, \textit{DMOS-E}, and \textit{DMOS-C}, each participant rated on a scale from 1 to 5, where 1 is Bad, 2 is Poor, 3 is Fair, 4 is Good, and 5 is Excellent. In each subjective experiment, the results from different methods in each evaluation group were randomly ordered. This randomization ensured that participants did not know which model or method produced each result, enabling a fairer assessment.

\subsection{Case Study}
\subsubsection{Examples of Understanding Empathetic Captions Using Chain-Talker}
To clearly demonstrate Chain-Taker's ability to understand and generate empathetic captions, Table~\ref{tab:exp-case-1} shows the captions produced by the model at different training epochs (50, 160, and 200), with the number of dialogue turns $N$ set to 3. Various stylistic attributes and their values are highlighted with colored rectangles for comparison. In the early stages of training, the model's understanding is inaccurate. For instance, at 50 epochs, the pitch is predicted incorrectly. As training epochs increase, the model gradually improves its ability to generate appropriate empathetic captions. By 200 epochs, the results are satisfactory. Compared to the Ground Truth, the model accurately predicts stylistic attributes such as pitch, speech rate, gender, and the emotion of anger. While it does not directly indicate the energy level, the phrase “is filled with a palpable sense of fury” also reflects the speaker's degree of anger.

\subsubsection{Examples of Generating Empathetic Captions Using CSS-EmCap}
In Fig.~\ref{fig:exp-case-2}, we show a set of dialogue data annotated with CSS-EmCap, which includes nine utterances. First, we identify the emotional categories and style attributes of each utterance. Then, we use the designed prompts to guide the LLM to generate basic descriptions and later to generate more suitable empathetic captions. The example demonstrates that the pipeline successfully detected the female speaker's negative emotion towards household chores (in dialogue turns 4 and 8). Additionally, we can see that the empathetic captions are more accurate and natural compared to the basic descriptions.

\end{document}